\documentstyle[12pt,epsf]{article}
\newcommand{\Del}{\Im m\Delta(\omega+i0^+)} 
\newcommand{\mDel}{-\Im m\Delta(\omega+i0^+)} 
\newcommand{\llangle}{\langle\!\langle}
\newcommand{\rrangle}{\rangle\!\rangle}

\begin{document}
\title{Numerical Renormalization Group Calculations for the Self-energy
of 
the impurity  Anderson  model}
\author{R Bulla$^1$,  A C Hewson$^2$ and Th Pruschke$^3$}
\date{}
\maketitle

{\it \noindent$^1$ Max-Planck-Institut f\"ur Physik komplexer
Systeme,          
 N\"othnitzer Str. 38,               
 01187 Dresden, Germany \\
$^2$ Department of Mathematics, Imperial College,
180 Queen's Gate, London SW7 2BZ, UK \\
$^3$ Institut f\"ur Theoretische Physik der Universit\"at,
93040 Regensburg, Germany}

\vskip .1in

\begin{abstract}
\noindent We present a new method to  calculate directly the one-particle 
self-energy
 of an
impurity Anderson model with Wilson's numerical Renormalization Group
method
by writing this quantity as the ratio of two correlation functions. This
way
of calculating $\Sigma(z)$ turns out to be considerably more reliable
and 
accurate
than via the impurity Green's function alone. We show results for the
  self-energy for the case of a constant coupling between impurity
  and conduction band ($\Del\!=\!$ const) 
  and the effective 
  $\Delta(z)$ arising in the Dynamical Mean Field Theory of the
Hubbard
  model. Implications to the problem of the metal-insulator 
  transition in the Hubbard model are also discussed.
\end{abstract}

\section{Introduction}
The single impurity Anderson model \cite{And61} is one of the most 
fundamental and probably the best understood model for strong electronic
correlations. Invented to describe the properties of magnetic impurities
in non-magnetic
metallic hosts 35 years ago, a variety of standard techniques have been
applied to it and new methods have been developed to study its static
and 
dynamic
properties in basically the whole parameter space (for a review see
e.g.\ 
\cite{Hew93}).
Although a very clear picture of the physics of the single impurity 
Anderson model has emerged from these calculations, a reliable method
for
calculating dynamic properties at very low temperatures and intermediate
or
large values of the Coulomb interaction was for a long time missing.

For example, Bethe ansatz calculations \cite{BA}, which are essentially
exact,
can only access static properties and the Quantum Monte Carlo method
\cite{Fye88}, which can be viewed as another numerically exact
technique, 
cannot
 reach very
low temperatures and/or large values of the Coulomb parameter, although
it
does not suffer from a minus sign problem here. In addition, the
analytic
continuation of the imaginary time data to real frequencies is a
numerically
highly ill-conditioned problem.

Among the approximate treatments the resolvent perturbation theory
together
with the so-called Non-Crossing Approximation
\cite{NCA} turned out to be a simple and powerful technique for high and 
intermediate
temperatures of the order of the Kondo scale but completely fails to
reproduce
the local Fermi-liquid properties as $T\!\to\!0$. Last but not least, 
straightforward
second-order perturbation theory in $U$ \cite{PT} has been shown to work 
surprisingly well down to
$T\!=\!0$ but it is
restricted to the symmetric case and not too large values of $U$.

The Numerical Renormalization Group method (NRG), invented by Wilson for
the
Kondo problem \cite{Wil75} and later applied by Krishnamurthy et al. to
the impurity Anderson model \cite{Kri80}, is usually acknowledged
primarily in
the context of universality and low-energy fixed point behaviour of the
Kondo
or Anderson model. One of its most appealing features is that it can
deal 
equally
well with small, intermediate or large values of $U$ and is not
restricted 
to half-filling.
During the last 15 years considerable progress has been made to extract 
dynamical properties with this method, too, and it has
been shown to give very accurate results also for e.g.\ dynamical one-
and 
two-particle and also transport properties \cite{Sak89,Cos94}.
The NRG works best at $T\!=\!0$, and various dynamic correlation
functions
can be calculated with an accuracy of a few percent. Although less well 
defined 
for finite temperatures, its extension to
$T\!>\!0$ also shows very good agreement with exact results 
\cite{Cos94}. It is quite remarkable
that no sum-rules (Friedel sum rule, total spectral weight) must be used
as
input
for these calculations. On the contrary, they can serve as an
independent
check on the quality of the results.

More recent interest in reliable methods to solve the impurity Anderson
model
and calculate its dynamical properties has been motivated by the 
discovery that
lattice
models in the limit of infinite dimensions acquire a purely local one
particle
self-energy \cite{Met89}. This simplification eventually leads to a
mapping of
the lattice problem on an effective impurity Anderson model coupled
to a medium to be determined self-consistently \cite{map}. Note that in
the
general case the solution of this self-consistency requires the
knowledge 
of the one 
particle self-energy.
In view of the wide
range of problems to which this so-called Dynamical Mean Field Theory
(DMFT,
see e.g.\ \cite{Geo96}) can be applied, it seems surprising why there
have 
been
hardly any contributions using the NRG. The only NRG-calculation 
known to us is the work of Sakai et al. \cite{Sak94} where the symmetric 
Hubbard model
in the metallic regime was studied. In their paper, these authors point
out 
some
difficulties in the progress of iterating the NRG results with the DMFT 
equations,
which are largely related to the necessary broadening of the NRG spectra
(see
further below).

In this contribution we present a new method to calculate dynamic
properties 
for 
the
impurity Anderson model, namely by directly constructing the interaction
contribution to the self-energy as the
ratio of two correlation functions, 
$\Sigma^U_\sigma(z)\! =\! U F_\sigma(z)/G_\sigma(z)$, with 
$F_\sigma(z) \!= \langle\!\langle f_{\sigma} f^\dagger_{\bar{\sigma}} 
f_{\bar{\sigma}},f^\dagger_{\sigma} \rangle\!\rangle_z $ 
and $G_\sigma(z) \!= \langle\!\langle f_{\sigma},f^\dagger_{\sigma}
\rangle\!\rangle_z $
(see Section II).
Details of how to calculate the $F(z)$ are given in the appendix. In
section III we discuss results for
\begin{itemize}
 \item the standard case, where the coupling between impurity states and
metallic host, $\Del$, is constant,
 \item and the Hubbard model in $d\!=\!\infty$, where  $\Delta(z)$
       has to be determined self-consistently.
\end{itemize}
The Hubbard model is studied in the paramagnetic regime, at half-filling
and $T\!=\!0$. We discuss the properties of self-energy and local
density
of states both in the metallic and insulating regimes
and some preliminary results for the metal-insulator transition.

\newpage

\section{Calculation of the self-energy}
\noindent
\paragraph{Model and basic concepts}
\mbox{}

The impurity Anderson model is written in the form
\begin{eqnarray}
  H &=&   \sum_{\sigma} \varepsilon_{\rm f} f^\dagger_{\sigma}
                             f_{\sigma}
                 + U  f^\dagger_{\uparrow} f_{\uparrow}
                       f^\dagger_{\downarrow} f_{\downarrow}
                \nonumber \\
           &+& \sum_{k \sigma} \varepsilon_k c^\dagger_{k\sigma}
c_{k\sigma}
            +  \sum_{k \sigma} V_k
           \Big( f^\dagger_{\sigma} c_{k \sigma}
               +   c^\dagger_{k\sigma} f_{\sigma} \Big). 
    \label{eq:siam}
\end{eqnarray}
In the model (\ref{eq:siam}), $c_{k\sigma}^{(\dagger)}$ denote standard
annihilation
(creation) operators for band states with 
spin $\sigma$ and energy $\varepsilon_k$,
$f_{\sigma}^{(\dagger)}$
those for impurity states with spin $\sigma$ and energy 
$\varepsilon_{\rm f}$. The
Coulomb interaction for two electrons at the impurity site is given by
$U$ and
both subsystems are coupled via a hybridization
$V_k$, which we allow to be $k$-dependent here.

Our final goal is to calculate the one-particle Green's function 
\mbox{$G_\sigma
(z)
=\langle\!\langle f_\sigma,f^\dagger_\sigma\rangle\!\rangle_z$}, which
formally can be written as
\begin{equation}
G_\sigma(z)=\frac{1}{z-\varepsilon_{\rm f}-\Sigma_\sigma(z)}\;\;.
\label{eq:Gfofz}
\end{equation}
While this formal introduction of the one particle self-energy 
$\Sigma(z)$
is straight forward, the actual calculation of $G(z)$ or alternatively 
$\Sigma(z)$
usually is an extremely complicated problem.
In order to express the self-energy $\Sigma(z)$ by standard impurity
correlation functions, we make use of the equation of motion
\begin{equation}
   z \langle\!\langle  A,B \rangle\!\rangle_z  + \langle\!\langle
{\cal L}A,B \rangle\!\rangle_z = \langle[ A,B]_\eta
\rangle,
    \label{eq:eom}
\end{equation}
with ${\cal L}\cdot \equiv [H,\cdot]_-$ and $\eta=+$, if both $A$ {\em
and}\/
$B$ are fermionic operators, $\eta=-$ otherwise. 
The correlation functions are defined as 
$\langle\!\langle  A,B \rangle\!\rangle_z \!=\! i\int_0^\infty
e^{izt}\langle[A(t),B]_\eta\rangle $.
For $A=f_{\sigma}$ and 
$B=f^\dagger_{\sigma}$ we obtain the equation of motion for the
f-Green's function as
\begin{equation}
   (z-\varepsilon_{\rm f}) G_\sigma(z) - 
     U \langle\!\langle f_{\sigma} f^\dagger_{\bar{\sigma}}
f_{\bar{\sigma}},
      f^\dagger_{\sigma} \rangle\!\rangle_z \nonumber - \sum_k V_k
       \langle\!\langle c_{k\sigma}, f^\dagger_{\sigma}
\rangle\!\rangle_z  = 1 \ .
\label{eq:eom_f}
\end{equation}
The correlation function 
$\llangle c_{k\sigma}, f^\dagger_{\sigma} \rrangle_z $
is related to $G_\sigma (z)$ via eq.\ (\ref{eq:eom}) with
$A=c_{k\sigma}$ and 
$B=f^\dagger_{\sigma}$ through  
\begin{equation}
  (z-\varepsilon_k) \llangle c_{k\sigma}, f^\dagger_{\sigma} \rrangle_z
     -V_k G_\sigma (z) =0 \ .
\label{eq:eom_cf}
\end{equation}
The $U$-term does not enter this equation as the Coulomb interaction
only
acts on the impurity states. Together with (\ref{eq:eom_cf}) eq.\ 
(\ref{eq:eom_f}) has the form
\begin{equation}
   (z-\varepsilon_{\rm f}) G_\sigma(z) - 
     U F_\sigma(z) - \Delta(z)  G_\sigma(z)
         = 1,
\label{eq:eom_f2}
\end{equation}
where we have defined 
$F_\sigma(z) = \llangle f_{\sigma} f^\dagger_{\bar{\sigma}}
f_{\bar{\sigma}},
      f^\dagger_{\sigma} \rrangle_z $
and $\Delta(z) = \sum_k V_k^2 \frac{1}{z-\varepsilon_k}$.
The total self-energy $\Sigma_\sigma(z)$ for the single impurity
Anderson 
model
is 
thus given by
\begin{equation}
     \Sigma_\sigma(z)=\Delta(z)+\Sigma_\sigma^U(z)\;\;,
\end{equation}
where the nontrivial part due to the Coulomb correlations $\Sigma^U(z)$
is
obtained from
\begin{equation}
    \Sigma^U_\sigma(z) = U \frac{F_\sigma(z)}{G_\sigma(z)} ,
\label{eq:Sigma}
\end{equation}
For simplicity and since we are only interested in the paramagnetic
situation
for the time being the spin index will be dropped in the following.

Alternatively, the interaction part of the self-energy can of course
also be
calculated directly from eq.\ (\ref{eq:Gfofz}) using
\begin{equation}
    \Sigma^U(z) =G_0(z)^{-1}-G(z)^{-1}, \ \ {\rm with}\ \
G_0(z) = \frac{1}{z-\varepsilon_{\rm f}- \Delta(z)}.
     \label{eq:Sigma_inv}
\end{equation}
On a first glance, there seems to be no apparent reason to 
prefer
the more complicated equation (\ref{eq:Sigma}) over equation
(\ref{eq:Sigma_inv}).
In order to clarify the advantage of using eq.\ (\ref{eq:Sigma})
instead of eq.\ (\ref{eq:Sigma_inv})
for the calculation of $\Sigma^U(z)$ with the NRG we want to give a
brief 
description
of
how the spectral densities for $G(z)$ and $F(z)$ are calculated with
the NRG.

\noindent\paragraph{Technical details}\mbox{}

Within the NRG, the impurity Anderson  model
eq.\ (\ref{eq:siam}) is mapped onto a semi-infinite chain (see 
\cite{Wil75,Kri80}) which is diagonalized iteratively starting from
the uncoupled impurity.
At each iteration, the number of states
increases by a factor of 4 and after a certain number of iterations,
the basis kept for the next iteration has to be truncated.
The important point of the method is that the coupling between
consecutive
elements of the chain decreases
exponentially for increasing distance from the origin, so that with
increasing
chain length at each iteration basically
only the lowest lying states will be renormalized and such a truncation
is 
meaningful. 
The spectral functions at each iteration are calculated from the
corresponding
matrix elements, which are in turn related to those of the previous
iteration.
This procedure is well established for the one-particle density of
states
$A(\omega)=-\frac{1}{\pi}\Im mG(\omega+i0^+)$ \cite{Sak89,Cos94} and can
straight forwardly be extended to
$B(\omega)=-\frac{1}{\pi}\Im mF(\omega+i0^+)$. For details we refer the
reader to the appendix.
Due to the truncation of states, the spectral function for the whole
frequency
range has to be built up from the data of all the iterations.

The resulting spectral function is a set of $\delta$-functions
at frequencies $\omega_n$ with weights $g_n$
which are 
broadened on a logarithmic scale
as
\begin{equation}
  g_n\delta(\omega - \omega_n) \longrightarrow 
    g_n\frac{e^{-b_n^2/4}}{b_n \omega_n\sqrt{\pi}} \exp \left[ 
      -\frac{(\ln\omega - \ln \omega_n)^2}{b_n^2} \right] .
\label{gauss}
\end{equation} 
This form of broadening was also  used in \cite{Sak89} and 
\cite{Cos94} and is especially adapted to the exponential variation
in energies peculiar to the NRG.
The width $b_n$ is chosen as $b$ independent of $n$ and we use values
$0.3\le b\le 0.6$.

It is well known that with this scheme the NRG gives already quite
accurate
results for $G(z)$ \cite{Sak89,Cos94}. However, one might anticipate
some 
problems with the
calculation of $\Sigma^U(z)$ using eq.\ (\ref{eq:Sigma_inv}). The
function
$G_0(z)^{-1}$ is, of course, known exactly since $\Delta(z)$ is a given 
quantity. Building the difference between an exactly known and a
numerically
determined function is usually very susceptible to numerical errors, 
especially
in regions where the result becomes small. Since this is expected to
happen
close to the Fermi level, i.e.\ in the physically most relevant region,
one is
likely to run into problems there.

One naive attempt to reduce these kind of inconsistencies and numerical
errors
when building the difference in eq.\ (\ref{eq:Sigma_inv}) is to treat
$G_0(z)^{-1}$ and $G(z)^{-1}$ on the same level, that is to calculate
$G_0(z)^{-1}$ via the NRG as well by setting $U=0$. However, since
according 
to
the theory of error propagation in sums or differences the absolute
errors 
add,
one must expect this procedure to be also ill-conditioned. 
If both $G_0(z)$ and $G(z)$ are known exactly, the difference
$G_0(z)^{-1}\!-\!G(z)^{-1}$  always gives a negative imaginary part for
the self-energy as there would be a pole in  $G(z)^{-1}$ for every
pole in $G_0(z)^{-1}$ at the same energy with equal or larger residue. 
This is no longer guaranteed as soon as both $G_0(z)$ and $G(z)$ are
only known approximately, and one has to
use rather large values of the broadening parameter $b$ to avoid
unphysical
oscillations in $\Sigma^U(z)$. This broadening in turn leads to a strong
suppression of the high energy peaks because spectral weight is shifted
from the center of the peak to its tails (to the high energy side due to
eq.\  (\ref{gauss})).

For the calculation of $\Sigma^U(z)$ via eq.\ (\ref{eq:Sigma}) on the
other hand
we do not expect to face these kind of problems that severely. Again,
both 
quantities are
calculated on the same basis by broadening the NRG results with
(\ref{gauss}),
i.e.\ with the same systematic error. This time, however, we {\em
divide}\/ 
them
 by each
other, which means that only the {\em relative}\/ errors will be
propagated,
leading to a numerically much more stable procedure.

\begin{figure}[htb]
\epsfxsize=5.4in
\epsffile{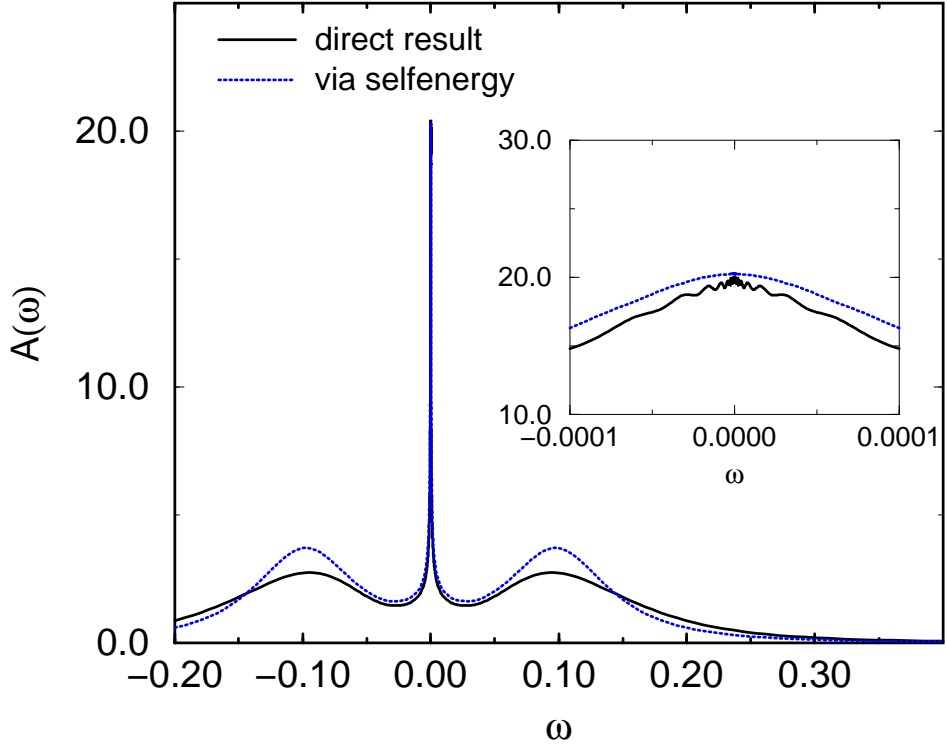}
\caption[]{Impurity spectral function for $\varepsilon_{\rm
f}\!=\!-0.1$,
$U\!=\!0.2$, $T\!=\!0$ and $\mDel=0.015$ in units
of the conduction electron band width.
$\bar{A}(\omega)$ (solid line) is the result  obtained directly from 
the NRG and
${A}(\omega)$ (dashed line) is calculated via the self-energy 
eq.\ (\ref{eq:Sigma}).
The inset shows the region around the Fermi-level.}
\label{fig:F1}
\end{figure}

Let us support this rather qualitative argument in  favour of expressing
the
self-energy as the ratio $UF(z)/G(z)$ by comparing the
spectral function $\bar{A}(\omega)$ obtained directly from the NRG
(solid line
in Fig.\ \ref{fig:F1}) and the ${A}(\omega)$ calculated from eq.\
(\ref{eq:Gfofz}) 
with the
self-energy expressed as in eq.\ (\ref{eq:Sigma}) (dashed line in Fig.\
\ref{fig:F1}). 
The spectral functions are defined as ${A}(\omega)=-\frac{1}{\pi}\Im m 
G(\omega+i0^+)$.
The parameters are $\varepsilon_{\rm f}\!=\!-0.1D$, $U\!=\!0.2D$ and
$\mDel
\!=\!\Delta_0\!=\!0.015D$, where $2D$ is the conduction electron
bandwidth.
For convenience we use $D\!=\!1$ as energy scale for the rest of the
paper.
The differences between the two methods can be summarized as follows:
\begin{itemize}
  \item We find for the total spectral weight 
        $\int d\omega\bar{A}(\omega)\!=\!\bar{w}\!=\!0.93$
        and $\int d\omega{A}(\omega)\!=\!{w}\!=\!0.9993$. The 7\%
        deviation in $\bar{w}$ can in principle be reduced by improving
        the
        resolution of the NRG calculation (smaller deviations
        have been achieved e.g.\ in \cite{Sak89,Cos94}). This is, however, not
        necessary
        in our case because the self-energy resulting from 
        eq.\ (\ref{eq:Sigma}) is an analytic function and the sum-rule 
        $w\!=\!1$ is then 
        automatically fulfilled (apart from the very small
        numerical error).
  \item The charge fluctuation peaks near $\varepsilon_{\rm f}$ are much
        more pronounced in $A(\omega)$. That the high energy features
        are usually underrated is a well-known problem in the
calculation of
        dynamical properties with the NRG. This problem is at least
partially
        resolved in our new scheme, since the main contribution in this
part
        of the spectrum rather comes from the hybridization part 
        $\Delta(z)$, which is treated 
        exactly.
  \item The oscillations in $\bar{A}(\omega)$ near $\omega\!=\!0$ 
        are due to the choice of the broadening which is obviously
        too small to see the correct low-frequency behaviour of
        the spectral function. These oscillations 
        almost vanish in ${A}(\omega)$ and the $\omega^2$-
        behaviour can be clearly identified.      
  \item The deviation from the Friedel sum-rule
        \begin{equation}
          A(0) = -\frac{1}{\pi\Im m\Delta(i0^+)}=:\frac{1}{\pi\Delta_0},
        \end{equation}
        ($\approx 21$ for the parameters used here)
        is about 7\% in $\bar{A}(\omega)$ and 4\% in ${A}(\omega)$.
\end{itemize}
Although the error in the Friedel sum-rule is visibly reduced, the
deviation 
is still
a few percent. Its origin will be discussed in the following.

\noindent\paragraph{Numerical aspects}\mbox{}

It is important to understand the origin of the deviation of $A(0)$ 
(Friedel sum-rule) from its  exact
value as this
lies at the heart of the numerical procedure.

Typical results for $A(\omega)$ and $B(\omega)$ as calculated with the
NRG
\begin{figure}[htb]
\epsfxsize=5.4in
\epsffile{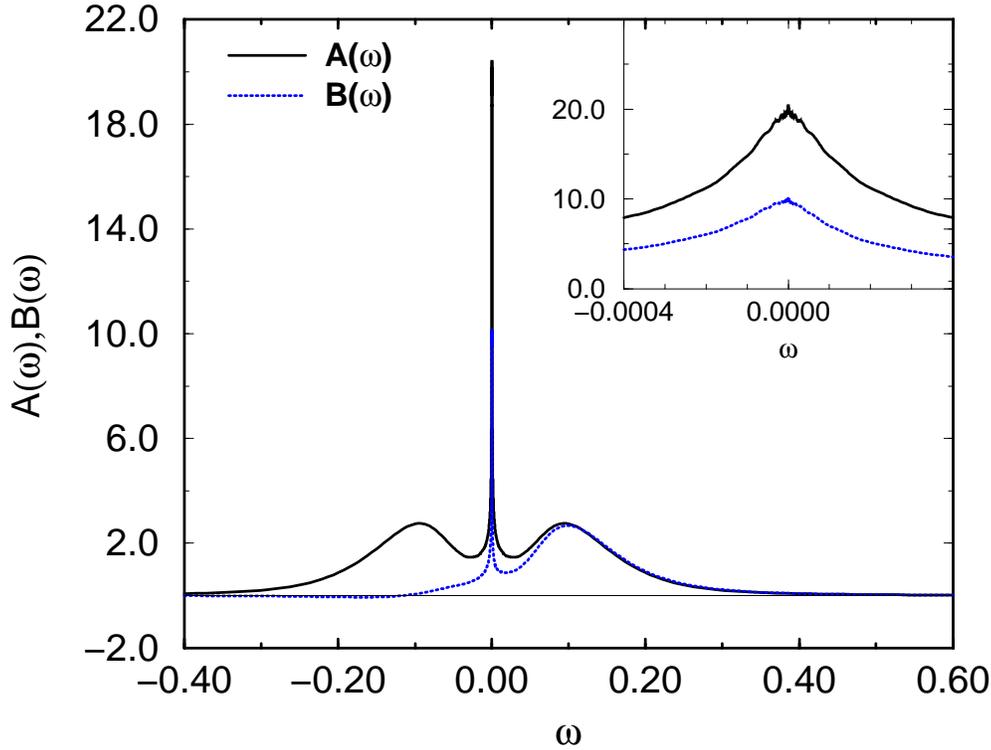}
\caption[]{Spectral functions $A(\omega)$ (solid line) and $B(\omega)$ 
(dotted line) for the same parameters as in Fig.\ 1
 (directly from the NRG, not via the self-energy).
The inset shows the region around the Fermi-level.}
\label{fig:F2}
\end{figure}
are shown in figure \ref{fig:F2}.
Both spectral functions $A(\omega)$ and $B(\omega)$ display a sharp
resonance
close to the Fermi energy. However, in contrast to $A(\omega)$, which is
positive definite and perfectly symmetric to $\omega=0$ due to the 
particle-hole
symmetry, the function $B(\omega)$ is obviously not positive definite
and
appears to be extremely asymmetric.

As next step we must calculate the real parts,
which are obtained via standard Kramers-Kronig transformation.
The self-energy finally is given with eq.\ (\ref{eq:Sigma}) as
\begin{equation}
  {\Re e}\Sigma^U(\omega+i0^+) + i{\Im m}\Sigma^U(\omega+i0^+) = 
  U \frac{{\Re e}F(\omega+i0^+) + i{\Im m}F(\omega+i0^+)}{{\Re e}
    G(\omega+i0^+) + i{\Im m}G(\omega+i0^+)} .
\end{equation}
In the particle-hole symmetric case, $\Re eG(\omega+i0^+)$ necessarily
vanishes
at $\omega\!=\!0$, therefore
\begin{equation}
  {\Re e}\Sigma^U(0^+) = U \frac{{\Im m}F(i0^+)}{
    {\Im m}G(i0^+)},
\end{equation}
which of course has to give the Hartree term $U/2$, and
\begin{equation}
  {\Im m}\Sigma^U(i0^+) = -U \frac{{\Re e}F(i0^+)}{
    {\Im m}G(i0^+)}.
\end{equation}
In the case of the standard single-impurity Anderson model we
furthermore know
that the Friedel sum-rule ${\Im m}\Sigma^U(i0^+)=0$ has to be fulfilled,
which
implies that
\begin{equation}
\Re eF(i0^+)=-\int\limits_{-\infty}^\infty
d\omega B(\omega){\cal P}\frac{1}{\omega}=0\;\;,
\label{eq:ReF0}
\end{equation}
where ${\cal P}(\ldots)$ denotes the principal value.
The relation (\ref{eq:ReF0}) is obviously not trivial regarding the
unusual
shape of $B(\omega)$. It indeed turns out that $\Re eF(i0^+)$ is
numerically zero
as long as the {\em full} spectrum of the Hamiltonian can be used.
However,
as soon as a truncation of states sets in the calculated value for $\Re
eF(i0^+)$
suddenly jumps to a finite value, eventually leading to a violation of
the
Friedel sum-rule as observed e.g.\ in Figure \ref{fig:F1}.

This observation suggests that also high energy states  are
important
to guarantee that $\Re eF(i0^+)\!=\!0$ and that a slight violation of the
Friedel 
sum-rule is almost unavoidable in this method.
\section{Results}
\noindent\paragraph{Single impurity Anderson model}\mbox{}

As a simple example let us discuss the standard case of a constant
$\Im m\Delta(\omega+i0^+)$,
\begin{equation}
 \mDel = \left\{ \begin{array}{lcl}
             \Delta_0 & : & |\omega|<1 \\
              0 & : & |\omega|>1
            \end{array}
     \right.   .
\end{equation}
\begin{figure}[htb]
\epsfxsize=5.4in
\epsffile{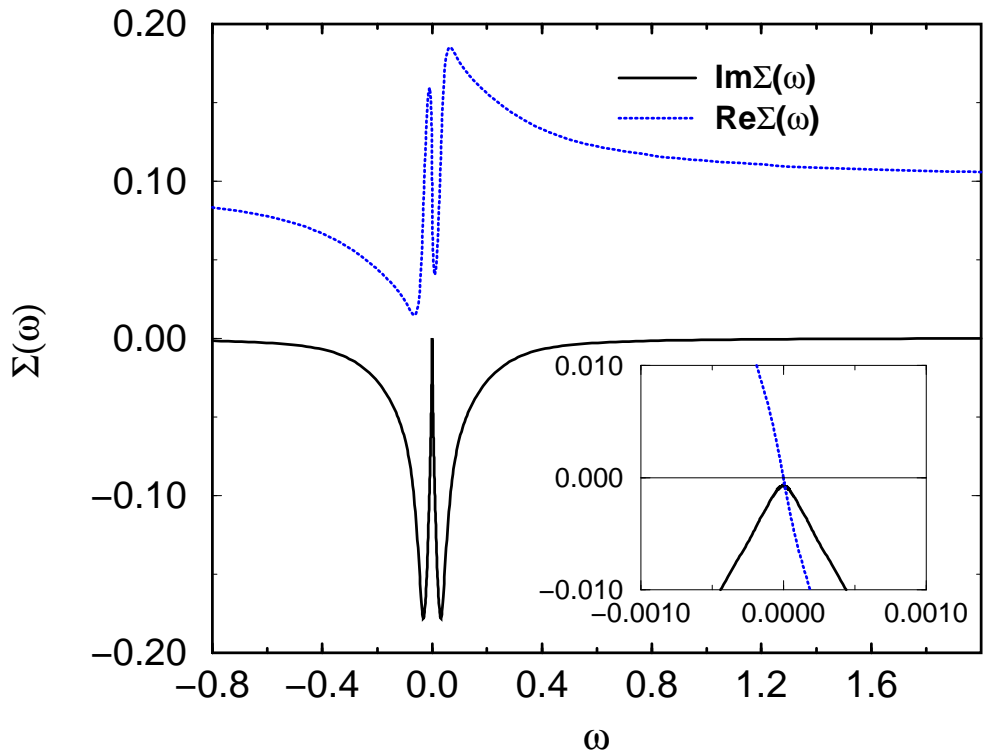}
\caption[]{Real and imaginary part of the self-energy for 
$\varepsilon_{\rm f}\!=\!-0.1$, $U\!=\!0.2$, and a constant
$\Delta_0\!=\!0.015$.
The inset shows the region around the Fermi level where the Hartree term
was subtracted off the real part.}
\label{fig:F3}
\end{figure}
The application of the NRG to this model has been discussed very
extensively
in the literature \cite{Sak89,Cos94}. Thus the results presented here
surely
give no new insight into the physics of this model. They are mainly
intended
to give the reader a feeling for the quality of our method.

Figure \ref{fig:F3} shows the results for the real and imaginary part of 
$\Sigma^U(z)$ for $\varepsilon_{\rm f}\!=\!-0.1$, $U\!=\!0.2$, 
$\Delta_0\!=\!0.015$ and $T\!=\!0$. 
As a first important point we note that the real part of the 
self-energy has a constant contribution. If we calculate $\Re
e\Sigma^U(\omega)+
\Re e\Sigma^U(-\omega)$, we obtain the expected value U/2
to within numerical precision for all $\omega$.
This result also shows that, although $B(\omega)$ is asymmetric, the
final
result obeys the particle-hole symmetry to a high precision.
In addition the slope 
$\left.\partial\Re e\Sigma^U(\omega)/\partial\omega\right|_{\omega=0}$
is negative and large, corresponding to a high effective mass.

The imaginary part of $\Sigma^U(z)$ shows two pronounced peaks at 
$\omega\!\approx\!\pm 0.03$ and a steep decrease as $\omega\to0$.
In the vicintiy of the Fermi level we find the Fermi liquid property
$\Im m\Sigma^U(\omega+i0^+)\!\propto\!\omega^2$ (inset
of Figure \ref{fig:F3}). However, as pointed out in the previous
section, 
the Friedel
sum rule $\Im m\Sigma^U(i0^+)=0$ is not exactly fulfilled. The shift
of $\Im m\Sigma^U(i0^+)\!\approx -0.0007$ corresponds to a 4\% error in
$A(0)\!=\!1/(\pi\Delta_0)$. 

\begin{figure}[htb]
\epsfxsize=5.4in
\epsffile{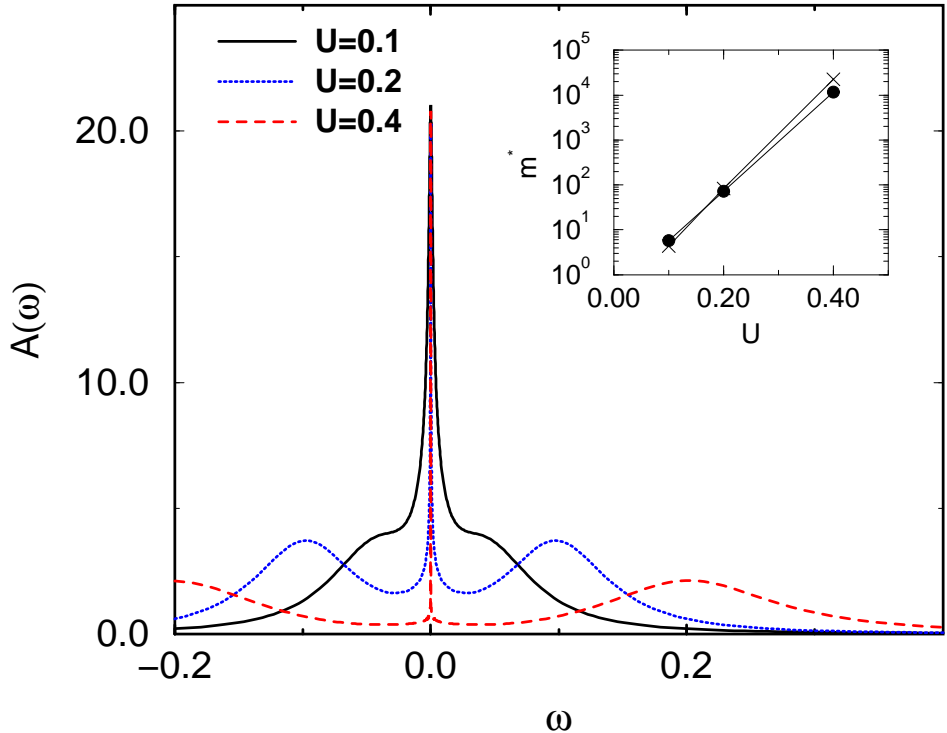}
\caption[]{Spectral function for 
$\varepsilon_{\rm f}\!=\!-U/2$,
$\Delta_0\!=\!0.015$ and various values of $U$. The inset shows the
resulting effective mass $m^\ast$ (filled circles) together with the
expected behaviour 
$m^\ast\propto \sqrt{U}\exp(\pi U/(8\Delta_0))$ (crosses).}
\label{fig:F4}
\end{figure}

Figure \ref{fig:F4} finally shows the resulting spectral function for 
various values of
$U$.
As mentioned previously in section 2, we find pronounced charge
fluctuation
peaks at $\pm U/2$ and the characteristic Abrikosov-Suhl resonance at
the 
Fermi
level. With increasing $U$ this resonance becomes sharper. The
corresponding
energy scale expressed via the effective mass is shown in the inset to
figure
\ref{fig:F4}
together with the expected behaviour 
$m^\ast\propto \sqrt{U}\exp(\pi U/(8\Delta_0))$.

\noindent\paragraph{Application to the Hubbard model}\mbox{}

The impurity Anderson model is not only useful to describe magnetic
impurities in nonmagnetic metals. It was shown only recently that in
the limit of infinite spatial dimensions a lattice model (Hubbard model,
Periodic Anderson Model, etc.) with local interactions can be mapped on
an
effective single impurity Anderson model.
The quantity $\Delta(z)$, which in the single impurity model describes
the coupling to the metallic host, becomes in general  an  energy
dependent
quantity here, which has a meaning similar to the Weiss field in the 
mean-field theory of the Heisenberg model. Since $\Delta(z)$ is a
dynamic
quantity which must be determined self-consistently
as functional of the one-particle self-energy \cite{Met89,map,Geo96}, the
name ``Dynamical Mean Field Theory'' (DMFT) has been coined.

This self-consistency makes it necessary to
calculate the self-energy $\Sigma^U(z)$ as accurate as possible.
Here we want to demonstrate that the NRG together with the method of
calculating $\Sigma^U(z)$ presented in the previous section is indeed
a reliable and accurate method to do this job at $T=0$.

The first step in order to apply the NRG is the mapping of the
impurity model on a semi-infinite chain for the case of a non-constant
$\Del$ which we have already described in \cite{BPH}.
As the resulting $\Del$ can develop very narrow structures at
the
Fermi level, we need a reliable numerical method to calculate $\approx$
60 - 100 hopping matrix elements of the chain. This is done
using arbitrary-precision fortran routines. 
Apart from the difference in the hopping matrix elements, 
the calculation of $F(z)$, 
$G(z)$ and $\Sigma^U(z)$ follows the same procedure as  in the flat-band
case.

The simplest model for correlation effects in solids is the well-known
Hubbard
model \cite{hub}.
This model is believed to have a rich phase diagram despite its
comparatively
simple form. DMFT studies at finite temperatures indeed revealed for
example
antiferromagnetic \cite{Jar95,Geo96} and ferromagnetic transitions 
\cite{Ulmke97,Pru97}, Mott-Hubbard metal
insulator transition \cite{Geo96} etc.\ . Nevertheless, there still
remain 
lots of
interesting open questions, especially about the properties of the model
at
extremely low temperatures both at and away from half filling.

\begin{figure}[htb]
\epsfxsize=4.4in
\epsffile{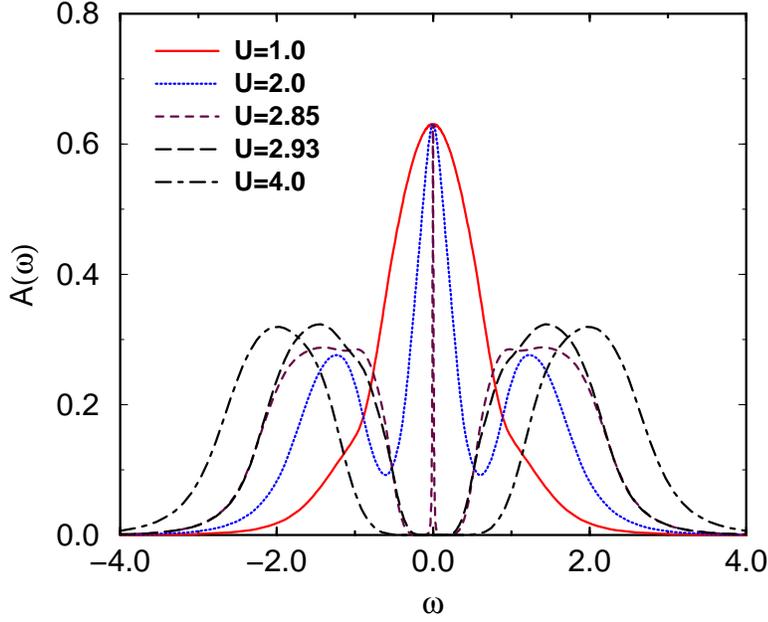}
\caption[]{Local spectral function of the Hubbard model for 
various values of $U$. A quasiparticle peak develops for increasing
values of $U$ which vanishes at a critical value $U_{\rm
c}\!\approx\!2.93$,
signalling the metal-insulator transition.}
\label{fig:spectra}
\end{figure}

Here we study the Hubbard model at $T=0$ for a semi-circular density of 
states $\rho_0(\varepsilon)$ corresponding to the
Bethe lattice with infinite coordination number
\begin{equation}
  \rho_0(\varepsilon) = \frac{2}{\pi}
\sqrt{1-\varepsilon^2}
  \label{eq:dos},
\end{equation}
($D\!=\!1$) at particle-hole symmetry and in the paramagnetic regime.
The resulting spectral functions for the paramagnetic Hubbard model for
various values of $U$ are collected in Figure \ref{fig:spectra}.
With increasing $U$, the one-particle spectrum develops the typical
three-peak
structure with a quasiparticle peak at $\omega\!=\!0$ and the two
Hubbard bands at $\pm U/2$.
Above a certain value $U_{\rm c}\approx 2.93$ 
the central peak vanishes and the system becomes insulating.

\begin{figure}[htb]
\epsfxsize=4.4in
\epsffile{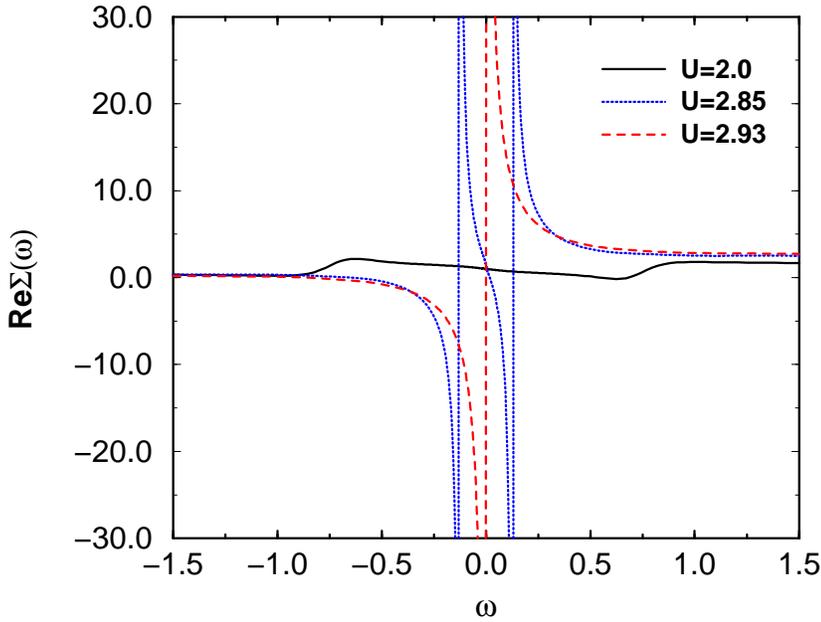}
\caption[]{Real part of the self-energy for the Hubbard model (same
parameters as in Fig.\ \ref{fig:spectra}). The negative slope at 
$\omega\!=\!0$ diverges
at the metal-insulator transition. For $U\!\ge\! U_{\rm c}$, the real
part shows a $1/\omega$-divergence.}
\label{fig:ReS}
\end{figure}

\begin{figure}[htb]
\epsfxsize=4.4in
\epsffile{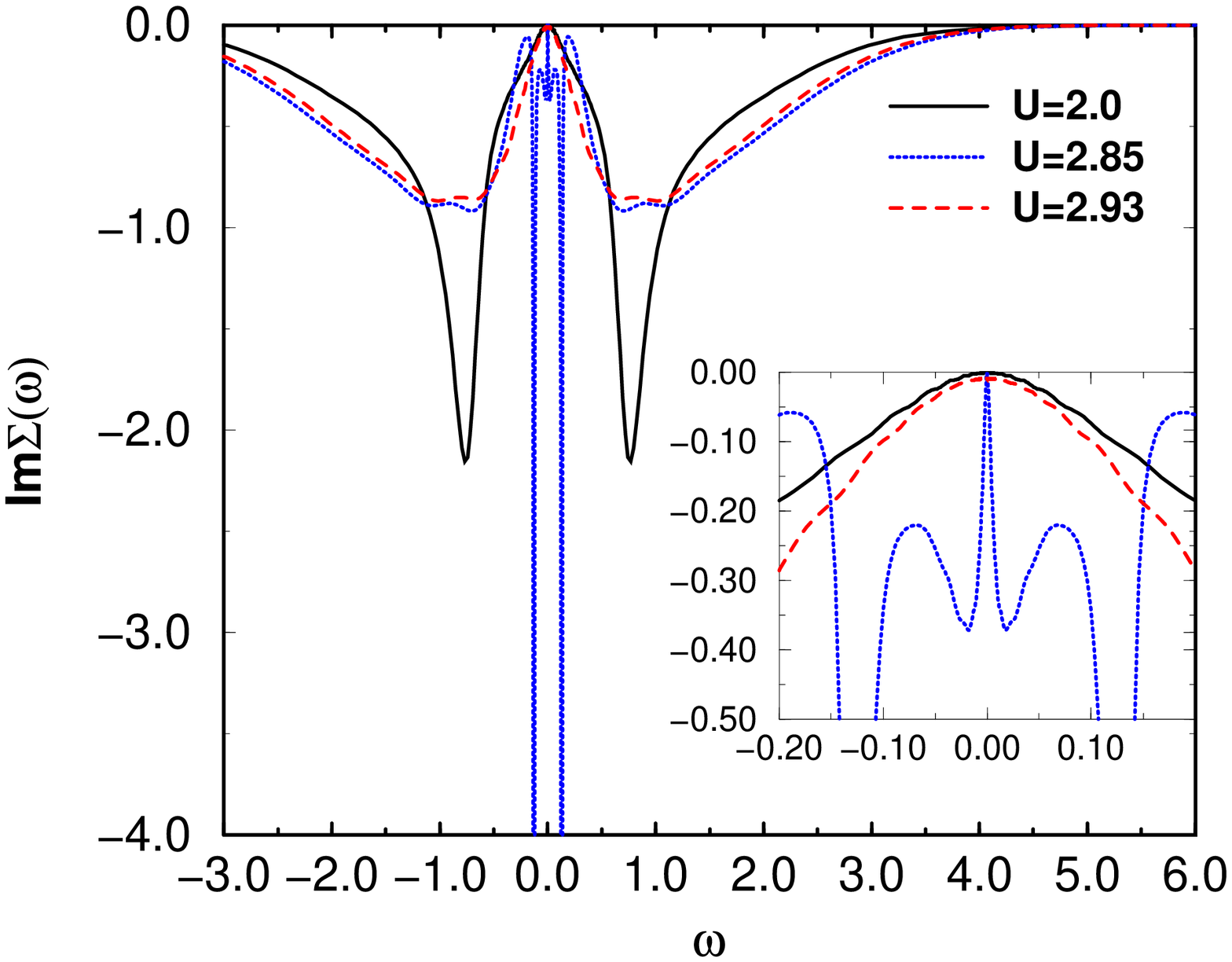}
\caption[]{Imaginary part of the self-energy for the Hubbard model (same
parameters as in Fig.\ 5). A $\delta$-function develops for $U\to 
U_{\rm c}$.}
\label{fig:ImS}
\end{figure}

Figure \ref{fig:ReS} and \ref{fig:ImS} show the real and imaginary part
of the self-energy
for the same parameters as in Fig.~\ref{fig:spectra} ($U\!=\!1$ and
$U\!=\!4$ are
not shown).
The Hartree term in the real part ($=\!U/2$) is subtracted.
The negative slope at the Fermi level diverges as $U\to U_{\rm c}$.
For $U\!\ge\!2.93$ (the insulating solution) the real part shows a
$1/\omega$-divergence. The corresponding $\delta$-peak in the
imaginary part is not plotted in Fig.\ \ref{fig:ImS}. This $\delta$-peak
in
Im$\Sigma^U(\omega)$ emerges from a two-peak structure in the metallic
regime, with the positions of the two peaks approaching $\omega\!=\!0$
for $U\to U_{\rm c}$. 
The imaginary part shows the Fermi liquid behaviour
Im$\Sigma^U(\omega)\propto \omega^2$ at low frequencies for $U< U_{\rm
c}$

\begin{figure}[htb]
\epsfxsize=5.8in
\epsffile{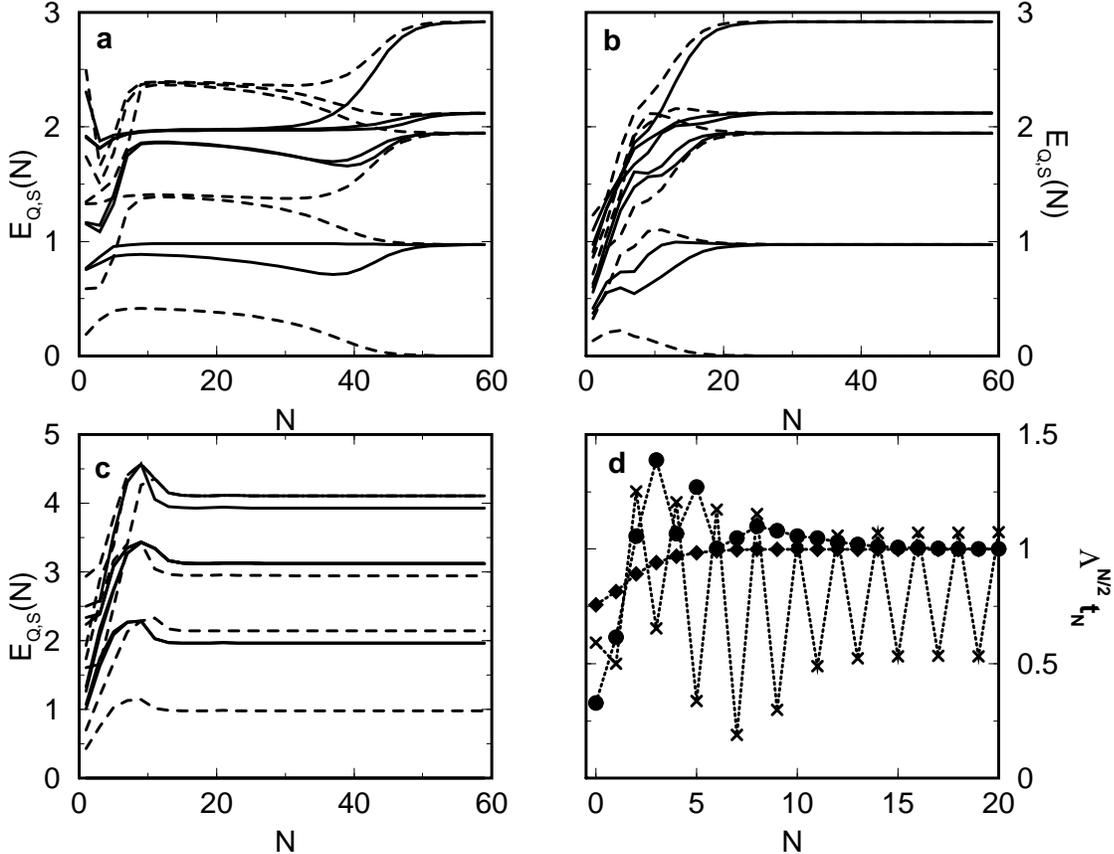}
\caption[]{Flow diagrams for the lowest energy levels $E_{Q,S}$ as
function of the NRG-iterations $N$.
The solid lines correspond to quantum numbers $Q\!=\!0,S\!=\!1/2$ and 
the dashed lines to quantum numbers $Q\!=\!1,S\!=\!0$.
(a) The flat-band case with $\varepsilon_{\rm f}\!=\!-0.2$, $U\!=\!0.4$,
and a constant $\Delta_0\!=\!0.015$. For large $N$, the system flows
to the Fermi liquid fixed point, while in the intermediate regime 
($N\approx20$)
it is near the so-called local moment fixed point. (b) The Hubbard model
with $U=2$ flows to the same Fermi liquid fixed point as in the flat band
case. (c) The Hubbard model with $U=4$ flows to the local moment
fixed point corresponding to the insulating behaviour.
(d) The hopping matrix elements $t_N$ of the semi-infinite NRG chain 
for Fig. \ref{fig:flow}a (diamonds), Fig. \ref{fig:flow}b (circles)
and Fig. \ref{fig:flow}c (crosses), respectively.}
\label{fig:flow}
\end{figure}

In order to give the reader an idea of the complex structures arising in
lattice models, we show in figure \ref{fig:flow} a comparison of the NRG
flow diagram
for the  
energy levels
for the single impurity Anderson model with flat $\Im m\Delta(\omega)$ 
(figure \ref{fig:flow}a)
and typical results for the Hubbard model in the paramagnetic metallic
phase (figure \ref{fig:flow}b) and paramagnetic insulating phase (figure
\ref{fig:flow}c). In contrast
to the single impurity case the flow diagrams for the Hubbard model show
a
complicated crossover behaviour for high energies (low NRG iteration
number)
before they saturate into a fixed-point spectrum for large 
NRG iteration numbers,
i.e.\ low energies. While these fixed-point spectra for the impurity
model and
the metallic solution of the Hubbard model (figures \ref{fig:flow}a and
b) are 
identical, 
i.e.\ both are Fermi liquids, the fixed point spectrum for the
insulator (figure \ref{fig:flow}c) has a quite different 
structure (see eg.\ the flow of the first excited state with 
$Q\!=\!1,S\!=\!0$ in figure \ref{fig:flow}b and figure \ref{fig:flow}c,
$Q$ is defined as the particle number with respect to half-filling).
In
addition the behaviour of the hopping matrix elements for the three
cases
is shown in figure \ref{fig:flow}d (diamonds, circles and crosses, 
respectively).
Note the oscillatory behaviour in the latter case characteristic for a
system
with a pseudogap.

The data shown here are quite similar to those obtained by Georges et 
al.\ \cite{Geo96} in that the quasiparticle peak seems to be isolated
from the two Hubbard bands near $U_{\rm c}$.
However, we always find finite spectral weight in the region
between the quasiparticle peak and the Hubbard bands.
For $U\ge U_{\rm c}$ the quasiparticle peak vanishes but we do not
see a real gap, i.e.\ a region with exactly vanishing density of
states. The fact that there is no gap even above the metal-insulator
transition
is also visible in $\Im m\Sigma^U(\omega)$, as there is no region with
vanishing  $\Im m\Sigma^U(\omega)$. The strong suppression of the
spectral
density 
between the Hubbard bands and the quasiparticle peak for $U\!=\!2.85$
mainly comes from the large values of $\Re e\Sigma^U(\omega)$.
The question, whether a real gap will eventually emerge for higher
values of 
$U$ is currently investigated and 
a more detailed analysis of the metal-insulator
transition at $T=0$ will
be presented in a subsequent publication.

For the time being we {\it define} the point where the
transition from a metal to an insulator takes place by the divergence
of the effective mass
\begin{equation}
  m^\ast=1-\left.\frac{\partial}{\partial \omega}{\Re e}
\Sigma^U(\omega)
  \right|_{\omega=0}
\end{equation}
of the quasi-particles. Note that 
this scenario completely neglects the possibility
of a discontinuous transition for a $U<U_{\rm c}$.

\begin{figure}[htb]
\epsfxsize=4.4in
\epsffile{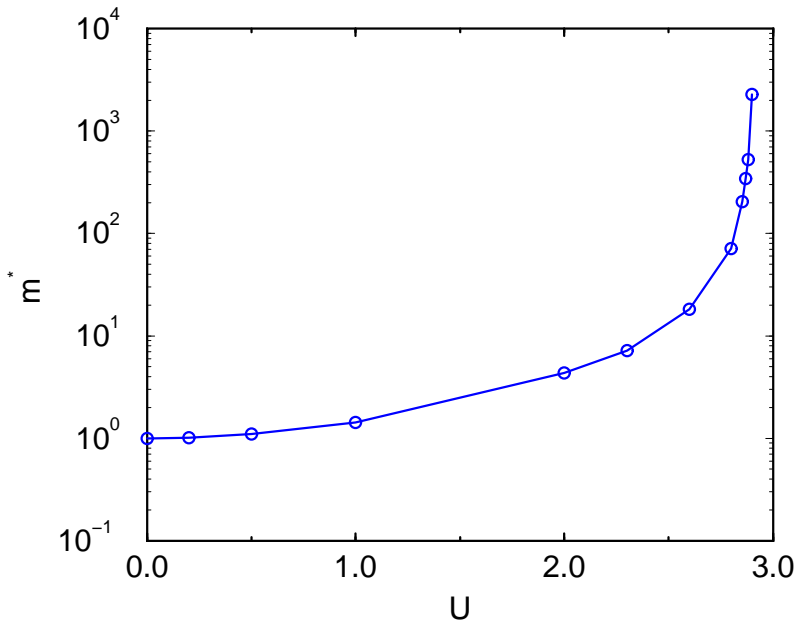}
\caption[]{$U$-dependence of the effective mass $m^\ast$ for the Hubbard
model. $m^\ast$ diverges at $U_{\rm c}\!\approx\!2.93$  which
defines the critical value of the metal-insulator transition.}
\label{fig:effmass}
\end{figure}

The behaviour of the effective mass as function of $U$  is collected in
figure \ref{fig:effmass}. $m^\ast$ diverges at $U_{\rm c}\!
\approx\!2.93$  and the critical behaviour close to $U_{\rm c}$ is
consistent with a powerlaw with an exponent of 
 $\approx -2$. Unfortunately, the data currently
available do not allow
a precise evaluation of this exponent.
Note that our value for  $U_{\rm
c}$
is considerably smaller than the value of $U_{\rm c,2}\!=\!3.3$
mentioned in the work of Georges et al.\ \cite{Geo96}. 

\section{Summary}

In this paper we have presented a new method of calculating the 
self-energy
of the single impurity Anderson model with the Numerical Renormalization
Group method. In contrast to the standard approach where one calculates
the self-energy from the Green's function alone, we express
$\Sigma^U(z)$
as a ratio of two correlation
functions.
The central aspect of this paper is that this method is much 
more accurate than the usual
method.

The importance of this gain in accuracy goes beyond the mere improvement
of the results for the single
impurity Anderson model.
Our method now in addition allows to apply the NRG to various lattice
models
within the Dynamical Mean Field Theory, where the self-energy of an
effective
impurity Anderson model has to be calculated self-consistently. 
As examples we recapitulated typical results for the single impurity
Anderson model and presented results for the
Hubbard model with a semi-circular density of states (corresponding to
the Bethe lattice with infinite coordination number) at particle-hole
symmetry and $T\!=\!0$. 

As an interesting but not yet fully confirmed
result in the latter case we find that
the metal-insulator transition in this model appears more like a 
metal-semimetal transition, as there is no region in the spectrum
where the spectral density exactly vanishes. We defined the
critical $U$ via the divergence of the effective mass and found
a value $U_{\rm c}\approx 2.93$.

A more detailed analysis of the metal-insulator transition,
results for the Hubbard model away from particle-hole symmetry 
and the investigation of more complicated models (Periodic Anderson
model, three-band Hubbard model, etc.)
will be discussed in forthcoming publications.

\vspace*{0.5cm}

We wish to thank T.\ Costi, J.\ Keller and D.\ Logan  
for a number of stimulating discussions. One of us (R.B.) was supported
by a grant from the Deutsche Forschungsgemeinschaft, grant No.\
Bu965-1/1.
We are also grateful to the EPSRC for the support of a research grant
(No.\ GR/J85349).

\appendix

\section{The correlation function $F(z)$}
Here we present some details of the calculation of
$F(z)$ and discuss some of its properties.

The NRG uses a discretized version of the Anderson model in a
semi-infinite
chain form (for details see \cite{Wil75,Kri80}). The resulting spectral
functions will therefore be given as a set of discrete $\delta$-peaks.

The spectral representation of $F(z)$ is 
\begin{eqnarray}
B(\omega) &=&  \frac{1}{Z} \sum_{nm} 
      <n|f_\downarrow f^\dagger_\uparrow f_\uparrow|m> 
        <m|f^\dagger_\downarrow|n>  \nonumber \\
    & & \times     \delta \left(  \omega -(E_m -E_n) \right)
     \left( e^{-\beta E_n} + e^{-\beta E_m} \right) .
 \label{eq:Fsum} 
\end{eqnarray}
The matrix elements $<n|f_\downarrow f^\dagger_\uparrow f_\uparrow|m> $,
 $<m|f^\dagger_\downarrow|n>$ 
and the energies $E_n, E_m$ are calculated iteratively in the NRG.
The two operators
\begin{eqnarray}
   V^{1/2}_{1/2} &=& f_\downarrow f^\dagger_\uparrow f_\uparrow
             , \nonumber \\
   V^{1/2}_{-1/2} &=& -f_\uparrow f^\dagger_\downarrow f_\downarrow ,
\end{eqnarray}
transform as 
\begin{eqnarray}
  \left[ s^\pm , V^{1/2}_{q} \right]_- &=&
      \sqrt{\frac{3}{4} - q(q\pm 1)} \ \ V^{1/2}_{q\pm 1} ,\nonumber \\
     \left[ s_z , V^{1/2}_{q} \right]_- &=&
      q V^{1/2}_{q} ,
\end{eqnarray}
($q=\pm 1/2$), with the spin operators
$$
   s^+ = f^\dagger_\uparrow f_\downarrow, \quad
   s^- = f^\dagger_\downarrow f_\uparrow, $$
\begin{equation}   s_z = \frac{1}{2}\left( f^\dagger_\uparrow f_\uparrow
              - f^\dagger_\downarrow f_\downarrow  \right).  
\end{equation}
This allows us to use the Wigner-Eckart Theorem 
\begin{equation}
         \left< Q,S,S_z,w \right\vert V^{1/2}_q
     \left\vert Q^\prime,S^\prime,S_z^\prime,w^\prime \right>
   = 
        \big< Q,S,w \big\vert \! \big\vert V_q^{1/2}
          \big\vert \! \big\vert Q^\prime,S^\prime,w^\prime  \big>       
         \big<S^\prime,S_z^\prime,\frac{1}{2},q\vert S,S_z\big>.
  \label{eq:WET}
\end{equation}
The $\big< Q,S,w \big\vert \! \big\vert V_q^{1/2}
          \big\vert \! \big\vert Q^\prime,S^\prime,w^\prime  \big>$
are reduced matrix elements and the 
$\big<S^\prime,S_z^\prime,\frac{1}{2},q\vert S,S_z\big>$
Clebsch-Gordan coefficients.
It is important to note that the operators $V_q^{1/2}$ transform in
exactly
the same way as the two operators
\begin{eqnarray}
   W^{1/2}_{1/2} &=&  f^\dagger_\uparrow ,
              \nonumber \\
   W^{1/2}_{-1/2} &=&  f^\dagger_\downarrow .
\end{eqnarray}
This has the consequence that all the recursion formulas for
the reduced matrix elements of $W^{1/2}_q$ can be used for the
calculation of the reduced matrix elements of $V^{1/2}_q$. The only
changes
are in the particle numbers $Q$ of the  states involved and the
initial values (see below).

The states $|n\!>$ and $|m\!>$ in eq.\ (\ref{eq:Fsum}) are classified in
terms
of charge $Q$ (the total particle number relative to the half-filled
case),
total spin $S$, $z$-component of the total spin $S_z$
and an additional label $w$
\begin{eqnarray}
    |n> &=& | Q_n,S_n,S_{z,n},w_n > \nonumber ,\\
    |m> &=& | Q_m,S_m,S_{z,m},w_m > \ .
\end{eqnarray}
The sum over $S_{z,n}$ and $S_{z,m}$ in eq.\ (\ref{eq:Fsum}) can be
performed exactly and we find
\begin{eqnarray}
   B(\omega) &=& \frac{1}{Z} 
    \sum_{\begin{array}{c} {\scriptscriptstyle Q,S,w_n} \\ 
           {\scriptscriptstyle S_m=S\pm\frac{1}{2},w_m}
       \end{array}}
   \big< Q,S,w_n \big\vert \! \big\vert V_{1/2}^{1/2} 
            \big\vert \! \big\vert
    Q+1,S_m,w_m \big>    \nonumber \\
    &\times& \big< Q+1,S_m,w_m \big\vert \! \big\vert
f_\downarrow^\dagger
            \big\vert \! \big\vert
    Q,S,w_n \big>   
        \delta \left(  \omega -(E_m -E_g) \right) \nonumber \\
    &\times& \left( e^{-\beta E_n} + e^{-\beta E_m} \right)
     \frac{1}{\sqrt{2}}\sqrt{2S+1} \nonumber \\
  &\times &  \left\{ \begin{array}{lcl}
             \sqrt{S} & : & S_m = S - \frac{1}{2} \\
             -\sqrt{S+1} & : & S_m = S + \frac{1}{2}
            \end{array}
     \right.  .    \label{eq:F_T}
\end{eqnarray}
We are only interested in the limit of zero temperature where we have
\begin{eqnarray}
   B^+(\omega) &=& \frac{1}{Z} 
    \sum_{S_m = S_g\pm \frac{1}{2}} \sum_{w_m}
   \big< Q_g,S_g,w_g \big\vert \! \big\vert V_{1/2}^{1/2} 
            \big\vert \! \big\vert
    Q_g+1,S_m,w_m \big>    \nonumber \\
    &\times& \big< Q_g+1,S_m,w_m \big\vert \! \big\vert
f_\downarrow^\dagger
            \big\vert \! \big\vert
    Q_g,S_g,w_g \big>   
        \delta \left(  \omega -(E_m -E_g) \right) \nonumber \\
    &\times& 
     \frac{1}{\sqrt{2}}\sqrt{2S_g+1} 
      \left\{ \begin{array}{lcl}
             \sqrt{S_g} & : & S_m = S_g - \frac{1}{2} \\
             -\sqrt{S_g+1} & : & S_m = S_g + \frac{1}{2}
            \end{array}
     \right. ,
\end{eqnarray}
for positive frequencies and
\begin{eqnarray}
   B^-(\omega) &=& \frac{1}{Z} 
    \sum_{S_n = S_g\pm \frac{1}{2}} \sum_{w_n}
   \big< Q_g-1,S_n,w_n \big\vert \! \big\vert V_{1/2}^{1/2} 
            \big\vert \! \big\vert
    Q_g,S_g,w_g \big>    \nonumber \\
    &\times& \big< Q_g,S_g,w_g \big\vert \! \big\vert
f_\downarrow^\dagger
            \big\vert \! \big\vert
    Q_g-1,S_n,w_n \big>   
        \delta \left(  \omega -(E_g -E_n) \right) \nonumber \\
    &\times& 
     \frac{1}{\sqrt{2}}\sqrt{2S_g+1} 
      \left\{ \begin{array}{lcl}
             -\sqrt{S_g} & : & S_n = S_g - \frac{1}{2} \\
              \sqrt{S_g+1} & : & S_n = S_g + \frac{1}{2}
            \end{array}
     \right. ,
\end{eqnarray}
for negative frequencies. The ground-state is labelled by
$|g>=\! | Q_g,S_g,S_{z,g},w_g >$, the ground-state energy is $E_g$ and
the partition function $Z$ reduces to the ground-state degeneracy.

To set up the iterative diagonalization of the reduced matrix elements\\
$\big< Q_n,S_n,w_n \big\vert \! \big\vert V_{1/2}^{1/2} 
            \big\vert \! \big\vert
    Q_m,S_m,w_m \big> $
we first of all need the initial values for the uncoupled impurity.
The only non-zero matrix element is
\begin{equation}
   \big<0,\frac{1}{2} \big\vert \! \big\vert V_{1/2}^{1/2} 
            \big\vert \! \big\vert 1,0 \big> = -1,
\end{equation}
in contrast to the two initial values
\begin{eqnarray}
   \big<0,\frac{1}{2} \big\vert \! \big\vert f_\downarrow^\dagger
            \big\vert \! \big\vert -1,0 \big> &=& 1 \nonumber, \\
   \big<1,0 \big\vert \! \big\vert f_\downarrow^\dagger
            \big\vert \! \big\vert 0,\frac{1}{2} \big> &=& -\sqrt{2}. 
\end{eqnarray}
Apart from the difference in the initial values and the
fact that $Q_m \!= \!Q_n+1$ for the 
$<\big\vert \! \big\vert V_{1/2}^{1/2}\big\vert \! \big\vert>$ 
matrix elements,
the recursion relations for both reduced matrix elements are identical
and
given by
\begin{eqnarray}
      \phantom{>}_N   \big< Q,S,w \big\vert\! \big\vert V_{1/2}^{1/2}
      \big\vert\!\big\vert
   Q^\prime,S^\prime,w^\prime \big>_N  &=& \sum_{rr^\prime}
   \sum_{pp^\prime=1}^4 U_{QS} (w,rp) U_{Q^\prime S^\prime}
(w^\prime,r^\prime
      p^\prime )                 \nonumber \\
     &\cdot &
         \phantom{>}_N   \big< Q,S,r;p \big\vert\! \big\vert
V_{1/2}^{1/2}
      \big\vert\!\big\vert
   Q^\prime,S^\prime,r^\prime;p^\prime \big>_N  \label{eq:app_rme} ,
\end{eqnarray}
with $p,p^\prime \in \{ 1,2,3,4 \}$. The $U_{Q,S}$ are the unitary
matrices
which diagonalize the Hamiltonian matrix in the subspace with
ccharge $Q$ and spin $S$.
The reduced matrix elements on the
right hand side of eq.\ (\ref{eq:app_rme}) are given by
\begin{eqnarray}
  _N   \big< Q,S,r;1 \big\vert \! \big\vert V_{1/2}^{1/2}
      \big\vert\!\big\vert
   Q+1,S\pm \frac{1}{2},r^\prime;1 \big>_N  &=&
     _{N-1}   \big< Q+1,S,r \big\vert \! \big\vert V_{1/2}^{1/2}
      \big\vert\!\big\vert
   Q+2,S\pm \frac{1}{2},r^\prime \big>_{N-1}  \nonumber \\
  _N   \big< Q,S,r;2 \big\vert \! \big\vert V_{1/2}^{1/2}
      \big\vert\!\big\vert
   Q+1,S + \frac{1}{2},r^\prime;2 \big>_N  &=&
      -\frac{2\sqrt{S^2 +S}}{2S+1}   
    _{N-1}   \big< Q,S-\frac{1}{2},r \big\vert \! \big\vert
V_{1/2}^{1/2}
      \big\vert\!\big\vert
   Q+1,S,r^\prime \big>_{N-1}  \nonumber \\
  _N   \big< Q,S,r;2 \big\vert \! \big\vert V_{1/2}^{1/2}
      \big\vert\!\big\vert
   Q+1,S - \frac{1}{2},r^\prime;2 \big>_N  &=&
      -
    _{N-1}   \big< Q,S-\frac{1}{2},r \big\vert \! \big\vert
V_{1/2}^{1/2}
      \big\vert\!\big\vert
   Q+1,S-1,r^\prime \big>_{N-1}  \nonumber \\
 _N   \big< Q,S,r;3 \big\vert \! \big\vert V_{1/2}^{1/2}
      \big\vert\!\big\vert
   Q+1,S + \frac{1}{2},r^\prime;3 \big>_N  &=&
      -
    _{N-1}   \big< Q,S+\frac{1}{2},r \big\vert \! \big\vert
V_{1/2}^{1/2}
      \big\vert\!\big\vert
   Q+1,S+1,r^\prime \big>_{N-1}  \nonumber \\
_N   \big< Q,S,r;3 \big\vert \! \big\vert V_{1/2}^{1/2}
      \big\vert\!\big\vert
   Q+1,S - \frac{1}{2},r^\prime;3 \big>_N  &=&
      -\frac{2\sqrt{S^2 +S}}{2S+1}   
    _{N-1}   \big< Q,S+\frac{1}{2},r \big\vert \! \big\vert
V_{1/2}^{1/2}
      \big\vert\!\big\vert
   Q+1,S,r^\prime \big>_{N-1}  \nonumber \\
_N   \big< Q,S,r;2 \big\vert \! \big\vert V_{1/2}^{1/2}
      \big\vert\!\big\vert
   Q+1,S - \frac{1}{2},r^\prime;3 \big>_N  &=&
      -\frac{1}{2S+1}   
    _{N-1}   \big< Q,S-\frac{1}{2},r \big\vert \! \big\vert
V_{1/2}^{1/2}
      \big\vert\!\big\vert
   Q+1,S,r^\prime \big>_{N-1}  \nonumber \\
_N   \big< Q,S,r;3 \big\vert \! \big\vert V_{1/2}^{1/2}
      \big\vert\!\big\vert
   Q+1,S + \frac{1}{2},r^\prime;2 \big>_N  &=&
      \frac{1}{2S+1}    
    _{N-1}   \big< Q,S+\frac{1}{2},r \big\vert \! \big\vert
V_{1/2}^{1/2}
      \big\vert\!\big\vert
   Q+1,S,r^\prime \big>_{N-1}  \nonumber \\
_N   \big< Q,S,r;4 \big\vert \! \big\vert V_{1/2}^{1/2}
      \big\vert\!\big\vert
   Q+1,S \pm \frac{1}{2},r^\prime;4 \big>_N  &=&
    _{N-1}   \big< Q-1,S,r \big\vert \! \big\vert V_{1/2}^{1/2}
      \big\vert\!\big\vert
   Q,S\pm \frac{1}{2},r^\prime \big>_{N-1}      \nonumber \\
\label{eq:app_rel}
\end{eqnarray}

The  spectral function $B(\omega)$ obeys the sum-rule
\begin{equation}
\int_{-\infty}^\infty {\rm d} \omega B(\omega) =  \frac{1}{Z} \sum_n
e^{-\beta E_n}
      <n| f^\dagger_\uparrow f_\uparrow|n> \equiv
        < f^\dagger_\uparrow f_\uparrow >,
 \label{eq:Fsum-rule} 
\end{equation}
which can be easily derived by integrating eq.\ (\ref{eq:Fsum}) over
$\omega$.
In the particle-hole symmetric case this gives
\begin{equation}
\int_{-\infty}^\infty {\rm d} \omega B(\omega) =  \frac{1}{2},
\end{equation}
where we also find the following relation between $B(\omega)$ and
$A(\omega)$
\begin{equation}
B(\omega) + B(-\omega) = A(\omega).  \label{eq:F_and_A}
\end{equation}
This can be directly obtained from
eq.\ (\ref{eq:eom_f2}).

\end{document}